\documentclass{sigchi}

\pdfoutput=1

\toappear{}
\usepackage{balance}  % to better equalize the last page
\usepackage{graphics} % for EPS, load graphicx instead
\usepackage{times}    % comment if you want LaTeX's default font
\usepackage{url}      % llt: nicely formatted URLs
\usepackage{subfig}

\makeatletter
\def\url@leostyle{%
  \@ifundefined{selectfont}{\def\UrlFont{\sf}}{\def\UrlFont{\small\bf\ttfamily}}}
\makeatother
\def\pprw{8.5in}
\def\pprh{11in}

\begin{document}

\title{Co-adaptation in a Handwriting Recognition System}

\numberofauthors{2}
\author{
  \alignauthor Sunsern Cheamanunkul\\
    \affaddr{University of California, San Diego}\\
    \affaddr{9500 Gilman Dr, La Jolla, CA 92093}\\
    \email{scheaman@eng.ucsd.edu}\\
  \alignauthor Yoav Freund\\
    \affaddr{University of California, San Diego}\\
    \affaddr{9500 Gilman Dr, La Jolla, CA 92093}\\
    \email{yfreund@eng.ucsd.edu}\\
}

% Teaser figure can go here
%\teaser{
%  \centering
%  \includegraphics{Figure1}
%  \caption{Teaser Image}
%  \label{fig:teaser}
%}

\maketitle

\begin{abstract}
  Handwriting is a natural and versatile method for human-computer
  interaction, especially on small mobile devices such as smart
  phones. However, as handwriting varies significantly from person to
  person, it is difficult to design handwriting recognizers that
  perform well for all users. A natural solution is to use machine
  learning to adapt the recognizer to the user. One complicating
  factor is that, as the computer adapts to the user, the user also
  adapts to the computer and probably changes their handwriting. This
  paper investigates the dynamics of co-adaptation, a process in which
  both the computer and the user are adapting their behaviors in order
  to improve the speed and accuracy of the communication through
  handwriting. We devised an information-theoretic framework for
  quantifying the efficiency of a handwriting system where the system
  includes both the user and the computer. Using this framework, we
  analyzed data collected from an adaptive handwriting recognition
  system and characterized the impact of machine adaptation and of
  human adaptation. We found that both machine adaptation and human
  adaptation have significant impact on the input rate and must be
  considered together in order to improve the efficiency of the system
  as a whole.
\end{abstract}


%\category{H.5.2}{Information Interfaces and Presentation (e.g. HCI)}{User Interfaces
%\textcolor{red}{See: \url{http://www.acm.org/about/class/1998/}
%for more information and the full list of ACM classifiers and descriptors. 
%Mandatory section: On the submission page
%only the classifiers' letter-number combination will need to be entered.}
%}

%\terms{
%	Algorithms; Measurement; Human Factors
%	If you choose more than one ACM General Term, 
%	separate the terms with a semi-colon.
%\\
%\textcolor{red}{If you choose more than one ACM General Term, 
%separate the terms with a semi-colon. See list of ACM terms at:
%\url{http://www.sheridanprinting.com/sigchi/generalterms.htm}.
%Optional section to be included in your final version.}
%}

\newtheorem{theorem}{Theorem}   
\newtheorem{lemma}[theorem]{Lemma}

% Force a column break
\vfill

\section{Introduction}
Handwriting is a natural and versatile method for human-computer
interaction, especially on small mobile devices such as smart
phones. As handwriting varies significantly from person to person, it
is difficult to design a handwriting recognition system that performs
well for all users. Modern handwriting recognizers resort to machine
learning techniques to adapt and specialize their handwriting models to
each individual user. As the recognizer adapts to the human user, the
user is likely to adapt to the system as well. We call this
situation ``co-adaptation'' where both human and computer adapts to
each other simultaneously.

In general, co-adaptation can manifest in any adaptive system.
Designing a system that co-adapts with the users is a challenging
problem on its own~\cite{Hook2000, Maes1994, Lim2009a}.  Our goal in
this paper is not to address those challenges, but rather to focus on
characterizing the impact of machine adaptation and of human
adaptation in the context of handwriting recognition. We believe that
this study will provide us with useful insights towards designing a
more efficient adaptive handwriting recognition system.

In order to evaluate performance of a handwriting recognition system
under co-adaptation, we introduce a framework based on the idea of
Shannon's communication channel~\cite{Shannon1948} that considers both
the user and the handwriting recognizer in a single system. Under this
framework, we define the notion of ``channel rate'' that measures the
amount of information successfully transfered from the user to the
computer. 

To quantify the effect of machine adaptation and of user adaptation
empirically, we developed a handwriting recognition system that is
capable of adapting to the handwriting of each individual user over
time. We collected usage data from 15 different users and performed an
analysis of the channel rate.

The paper is organized as follows. First, in
Section~\ref{sec:channel}, we present the information-theoretic
framework for quantifying the efficiency of a handwriting system where
the system includes both the user and the computer. Next, in
Section~\ref{sec:recognition_algorithm}, we describe our adaptive
handwriting recognition algorithm that we developed for our
experiment. Then, in Section~\ref{sec:experiment}, we describe the
experiment and present the results in terms of the performance
measures derived from the proposed framework. Finally, we draw some
conclusions in Section~\ref{sec:conclusions}.

\section{Handwriting recognition as a communication channel}
\label{sec:channel}
Unlike typing, which transmits information to the computer at discrete
time points, handwriting continuously transmits information as the
writer creates the trajectory. Traditionally, handwriting data is
analyzed one ``unit'' at a time where ``unit'' can be a stroke, a
character, a word or even a sentence. In this work, we propose an
alternative analysis where the data is analyzed in fixed intervals of
time. We consider the process of writing as a process through which
the intended letter is disambiguated from the other possible letters.

% input
\newcommand{\intent}{M}
\newcommand{\intentSet}{\mathcal{E}}
\newcommand{\intentDist}{\mathcal{M}}
% output
\newcommand{\pred}[1]{\mathcal{Q}_{#1}}
\newcommand{\predFinal}{\pred{\mathrm{final}}}
% trajectory
\newcommand{\writing}[1]{W_{1:{#1}}}
\newcommand{\writingVec}{\overline{W}}
\newcommand{\writingDist}{P(\writingVec | \intent)}
% misc
\newcommand{\tFinal}{T_{\mathrm{final}}}
\newcommand{\expectedDuration}{\mathbb{E} \left[\tFinal\right]}
\newcommand{\condEntropy}{H(\predFinal | \intent)}

We formalize this process using the concept of communication
channel~\cite{Shannon1948}.  Let $\intentSet$ denote the set of all
possible input. Technically, the set $\intentSet$ can be a set of
sentences, a set of words, or a set of characters. Without loss of
generality, in this work, we assume that $\intentSet$ is a set of 26
English characters. We also ignore dependencies between characters due
to the language model and due to the co-articulation effects between
neighboring handwritten characters.

\begin{figure*}[th]
  \centering
  \includegraphics[width=0.9\textwidth]{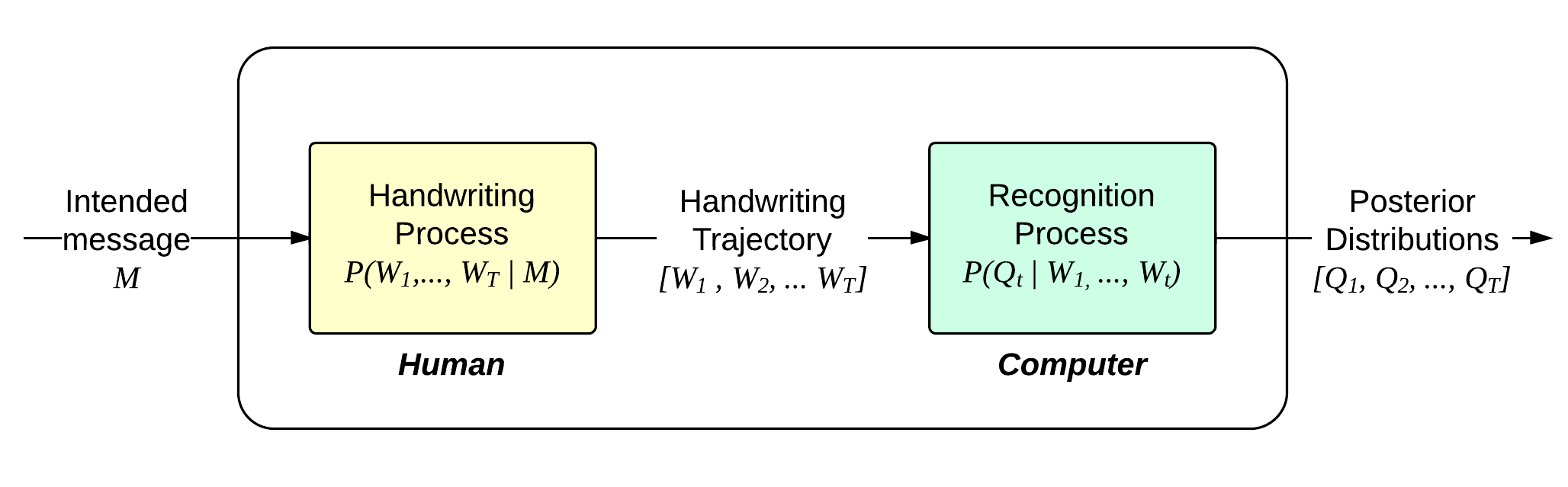}
  \caption{A summary of the handwriting recognition channel.}
  \label{fig:hwr_channel}
\end{figure*}

As shown in Figure~\ref{fig:hwr_channel}, the channel is comprised of
two separate processes. First, the handwriting process is the process
of which the user translates an intent $\intent \in \intentSet$ into a
series of hand movements which is sampled at some rate to create a
discrete time trajectory: $\writing{T} = \left[ (x_1,y_1), \ldots,
  (x_T,y_T) \right]$. In other words, this process {\em encodes} the
intent $\intent$ into a trajectory $\writing{T}$. Let $\writingVec$
denote the entire trajectory vector. The distribution $\writingDist$
denotes the variability of the encoding process. The second process is
the recognition process that decodes the handwriting trajectory back
into the original intent. For each time step $t$ where $1 \le t \le
T$, the process maps a trajectory $\writing{t}$ to a distribution over
$\intentSet$, denoted by $\pred{t}$.

Let $\tFinal$ and $\predFinal$ denote the final writing duration and the
posterior distribution when the user finishes writing the
trajectory. According to the theory of channel capacity, the
information transmitted through the channel can be quantified by the
mutual information between the input $\intent$ and the decoding
posterior $\predFinal$, denoted by $I(\intent ; \predFinal)$. We
define the mean posterior of $\predFinal$ conditioned on $\intent$ and
the average posterior distribution as follows.
\[
P(\predFinal | \intent) =
\int\limits_{\writingVec \sim \writingDist} { P(\predFinal | \writingVec)
P(\writingVec | \intent)} 
\]
\[
P(\predFinal)
=
\sum_{m \in \intentSet} 
P(\intent = m) P(\predFinal | \intent = m)
\]
Given these two expressions, we define the mutual information
between the character $\intent$ and the decoding $\predFinal$ as
\[
I(\intent ; \predFinal) = 
H(\predFinal)
- \sum_{m \in \intentSet} P(\intent = m) H(\predFinal | \intent = m)
\]
where the entropy of $\predFinal$ is defined as
\[
H(\predFinal) = -\sum_{m \in \intentSet} {
P(\predFinal = m) \log_2 P(\predFinal = m)}
\]

\newcommand{\RMI}{R_{\mathrm{MI}}}
\newcommand{\RLL}{R_{\mathrm{LL}}}

Next, we can define the channel rate in terms of the mutual
information and expected writing duration as 
\begin{align}
\label{eq:channel_rate}
\RMI
= 
\frac{I( \intent ;  \predFinal)}{\expectedDuration}
\end{align}
However, the channel rate $\RMI$ is not suitable for practical
implementation for two reasons. First, the estimation of $H(\predFinal
| \intent)$ requires an extensive amount of data. Secondly, suppose
the original intent is $m$, $\RMI$ yields a high value as long as
$P(\predFinal | \intent = m)$ concentrates any single intent $n$ even
when $n \ne m$.  Thus, we propose an alternative measure to the $\RMI$
based on the idea of log loss, called $\RLL$. We define $\RLL$ to be
\begin{align}
\resizebox{0.9\columnwidth}{!}{$
\label{eq:channel_rate_ll}
\RLL = \frac{H(\predFinal) - \displaystyle\sum_{m \in \intentSet}
  P(\intent = m) (-\log_2 P(\predFinal = m | \intent =
  m))}{\expectedDuration}
$}
\end{align}

The relationship between $\RMI$ and $\RLL$ is worth noting. When
$(-\log_2 P(\predFinal = m | \intent = m))$ is small then the
conditional entropy $H(\predFinal | \intent)$ is also small. As a
result, the mutual information $I( \intent ; \predFinal)$ will be
close to its maximal possible value of $H(\predFinal)$. In other
words, the log loss term $(-\log_2 P(\predFinal = m | \intent = m))$
provides an upper bound for the conditional entropy $H(\predFinal |
\intent)$ up to some constant factor. For the remaining of this paper,
when we refer to the {\em channel rate}, we strictly refer to $\RLL$.

Intuitively, the channel rate is a measure that quantifies both
accuracy and speed of a handwriting recognition channel at the same
time. Handwriting, as well as many other motor control tasks, obeys
the speed-accuracy tradeoff~\cite{Fitts1954}. It is not sufficient to
quantify the efficiency of a handwriting recognition system by its
recognition accuracy alone. For example, a system that requires the
user to write each character in a specialized form may attain a very
high recognition accuracy, but it would require the user more time and
effort to use. Such system might not be as efficient as a system that
makes more errors but allows the user to write freely. This leads us
to believe that the channel rate is a suitable measure that any
handwriting recognition system should aim to maximize. In a sense,
maximizing the channel rate is equivalent to finding a balance between
maximizing the recognition accuracy and minimizing the writing time
and effort of the user.

Based on this framework, it follows that the channel rate can be
improved by a combination of human learning and machine learning,
which corresponds to improving the handwriting process and the
recognition process respectively. Ideally, $\predFinal$ will always be
concentrated on the original intent $\intent$. This would mean that
the channel is perfect and works without error. However, in real-world
scenarios, errors will occur. One source of errors comes from mistakes
made in the recognition process. These recognition errors can be
reduced using training data and machine learning. The harder problem
is when there is a significant overlap between $\writingDist$ for
different intents. In this situation, we will need to rely on the user
to make their handwriting less ambiguous. Although the effect of human
learning is always present, we believe that it can be enhanced by
giving useful feedback to the user in the form of guidance or lessons.

\section{Adaptive recognition algorithm}
\label{sec:recognition_algorithm}

\newcommand{\prototypeSet}{\mathcal{P}} 

We developed an adaptive handwriting recognition algorithm that, for
every time step $t$, maps a partial handwriting trajectory
$\writing{t}$ to a posterior distribution over $\intentSet$, denoted
by $\predFinal$. By realizing that the effect of user adaptation is
likely to be present, we designed our recognition algorithm so that it
can adapt not only to each individual user, but also to the changes of
the handwriting trajectory distribution $\writingDist$ unique to each
user over time. The idea of specializing and adapting the recognizer
for each user has been studied and shown to be effective in reducing
the error rate~\cite{Connell2002, Matic93, Kienzle06}.

At a high-level, our adaptive recognition system can be outlined as
follows. For each user, the system creates and maintains one or more
character models for each character in $\intentSet$. We refer to each
of such models as a {\em prototype}. Each prototype is basically a
representative handwriting instance from the user. Technically, the
prototypes can be viewed as left-to-right hidden Markov models with
Gaussian observation~\cite{ThomasPloetz2011}. Let $\prototypeSet_u$
denote the set of prototypes for a user $u$. The adaptivity of our
system comes directly from the fact that $\prototypeSet_u$ is modified
over time. In the decoding process, given a handwriting trajectory and
a set of prototypes $\prototypeSet_u$, the system computes a posterior
distribution $\predFinal$ and, when a single prediction is needed, the
element with the maximum likelihood is predicted.

\subsection{Feature vectors and distance function}
In addition to the x- and y-coordinate, each handwriting trajectory is
supplemented with writing direction information. Specifically, each
handwriting instance is represented by a sequence of feature vectors
$\langle f_1, \ldots, f_T \rangle$ where $f_i = (x_i,y_i,
dx_i,dy_i)$. $(x_i,y_i)$ denotes the normalized touchscreen coordinate
and $(dx_i,dy_i) = (\frac{x_i - x_{i-1}}{z}, \frac{y_i - y_{i-1}}{z})
, z = \sqrt{(x_i - x_{i-1})^2 + (y_i - y_{i-1})^2}$ denotes the
writing direction.

To measure the similarity between two handwriting instances, we use
{\it dynamic time warping} (DTW) distance~\cite{Rabiner1993} as the
distance function in our algorithm. The DTW distance is commonly used
for variable-length data such as handwriting and speech. The
calculation can be done efficiently using dynamic programming.

\subsection{Initial adaptation}
The initial adaptation is critical for any intelligent system. It is
unquestionable that the performance of any well-behaved intelligent
system increases as the system learns more about the user. If the
initial adaptation is poor, the users might get frustrated with the
system and stop using it even before it can fully adapt to them.
 
We address the problem of initial adaptation by sharing data across
different users. Typically, people do have similar handwriting
especially when they share the same educational culture. The process
of the initial adaptation can be described as follows. In the very
first interaction with the user $u$, our system has no information
about the user and, therefore, assign a set of typical prototypes
which has been trained using data from multiple users in the
past. Specifically, the typical prototypes are the centroids of the
clusters returned by running a clustering algorithm (k-means) on a set of
training handwriting instances.  We refer to this set of prototypes as
$\prototypeSet_0$. After the first interaction, the system creates a
new set of prototypes $\prototypeSet_{(u,1)}$ by recomputing the
centroids of the clusters after adding the examples from the user to the
pool with significantly higher weights than the rest.

\subsection{Adapting the prototypes over time}
After collecting a few examples of the user's handwriting, the system
again performs the weighted clustering algorithm on the data to generate
a new set of prototypes $\prototypeSet_{(u,i+1)}$. In this stage, only
examples from the user and previous prototypes are considered.
This adaptation process happens after 3-5 new examples are acquired.

\begin{figure}[th]
  \centering
  \includegraphics[width=0.9\columnwidth] {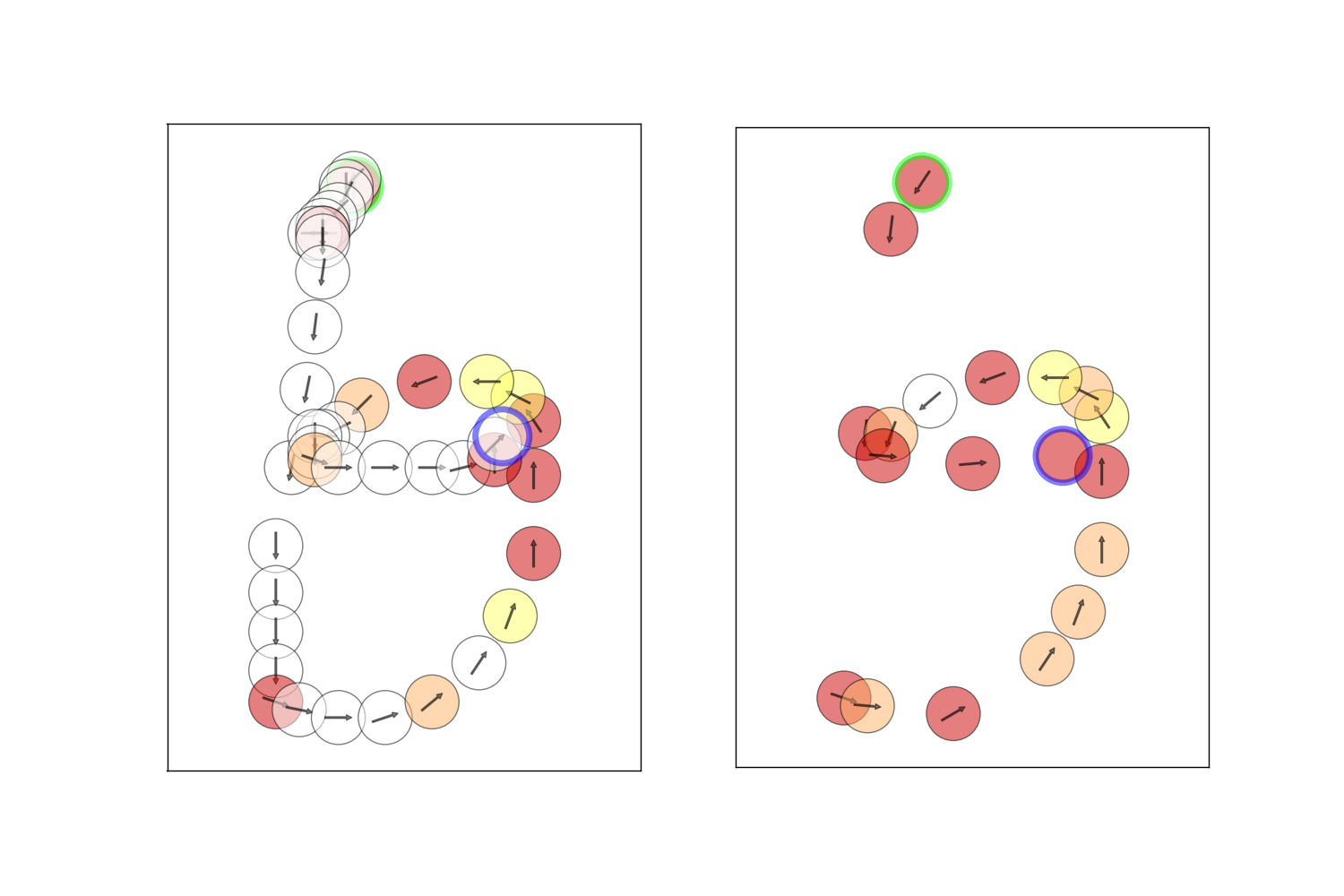}
  \caption{The hidden state reduction process is applied to each
    prototype to remove rarely visited states with respect to the
    training set. The originally trained prototype is shown on the
    left and the reduced prototype is shown on the right. The
    intensity of the colors corresponds to the expected number of
    times the state being mapped to. }
  \label{fig:state_reduction}
\end{figure}

To improve real-time performance, we need to keep the lengths (number
of states) of the prototypes as small as possible. After the new
prototypes are chosen, the system performs an additional step to
shorten the length of each prototype. This pruning process is similar
in spirit to removing and merging unnecessary hidden states in an HMM.
The basic idea is to remove unwanted states while maintaining the same
recognition power using a variant of forward-backward
algorithm~\cite{Bilmes97}. Figure~\ref{fig:state_reduction} shows the
hidden states before and after the reduction step.

\subsection{Decoding}
Our decoding algorithm is based on the standard Bayesian
inference. Namely, given a trajectory $\writing{T}$ and the current
set of prototypes $\prototypeSet_u$, the algorithm computes the
distance from $\writing{t}$ to each of the prototypes in
$\prototypeSet_u$ for all $1 \le t \le T$. The distances are then
transformed into a probability distribution $\pred{t}$. We use
$e^{-x}$ as the transfer function. When a single prediction is
expected, the algorithm simply returns the prediction with the maximum
likelihood.

\section{Experiment}
\label{sec:experiment}

\newcommand{\RAdapt}{R_{\mathrm{adapt}}}
\newcommand{\RFixed}{R_{\mathrm{fixed}}}
\newcommand{\RIdeal}{R_{\mathrm{ideal}}}

The main objective of our experiment is to determine and quantify the
effect of machine adaptation and of human adaptation when the users
interact with the system over some period of time. We implemented the
handwriting recognition system described in
Section~\ref{sec:recognition_algorithm} as an application on Apple iOS
platform. The application was presented to the users as a writing
game. In each session, each participant was presented with a random
permutation of the 26 lowercase English alphabets i.e. $\intentSet =
\left[ a \ldots z \right]$ and $P(\intent)$ is uniform. The objective
of the game was to write the presented characters as quickly as
possible and, more importantly, the handwritten characters should be
recognizable by the system. A score, which is the average {\it channel
  rate} of the session, was given to the user right after each session
to reflect the performance of the session. There were 15 participants
in this experiment. We asked them to play our game for at least 20
sessions over multiple days in his/her own pace. We did not control
past experience of the participants. Some of them had more experience
with touch screens than others.

The experiment was set up to demonstrate a condition called {\em
  co-adaptation} where both the user and the computer were allowed to
adapt together. We denote this condition $\RAdapt$. To investigate the
effect of co-adaptation, we create a controlled condition called
$\RFixed$ where the computer was not allowed to adapt with the
user. In other words, we ran a simulation to figure out what the
channel rates would have been if the prototype sets were never changed
from $\prototypeSet_0$. Ideally, it would be more preferable to have
$\RFixed$ determined by another control group where the prototypes
were kept fixed and never changed. However, the results from the
simulated condition can be seen as a lower bound on the amount of the
improvement attributable to human learning and, therefore, it is
sufficient to demonstrate our point.

\section{Results and discussion}
\label{sec:results}
 
\begin{figure}[tb]
  \centering
  \subfloat[$\RAdapt$]{
    \label{fig:channel_rate_adapt}
    \includegraphics[width=0.75\columnwidth]{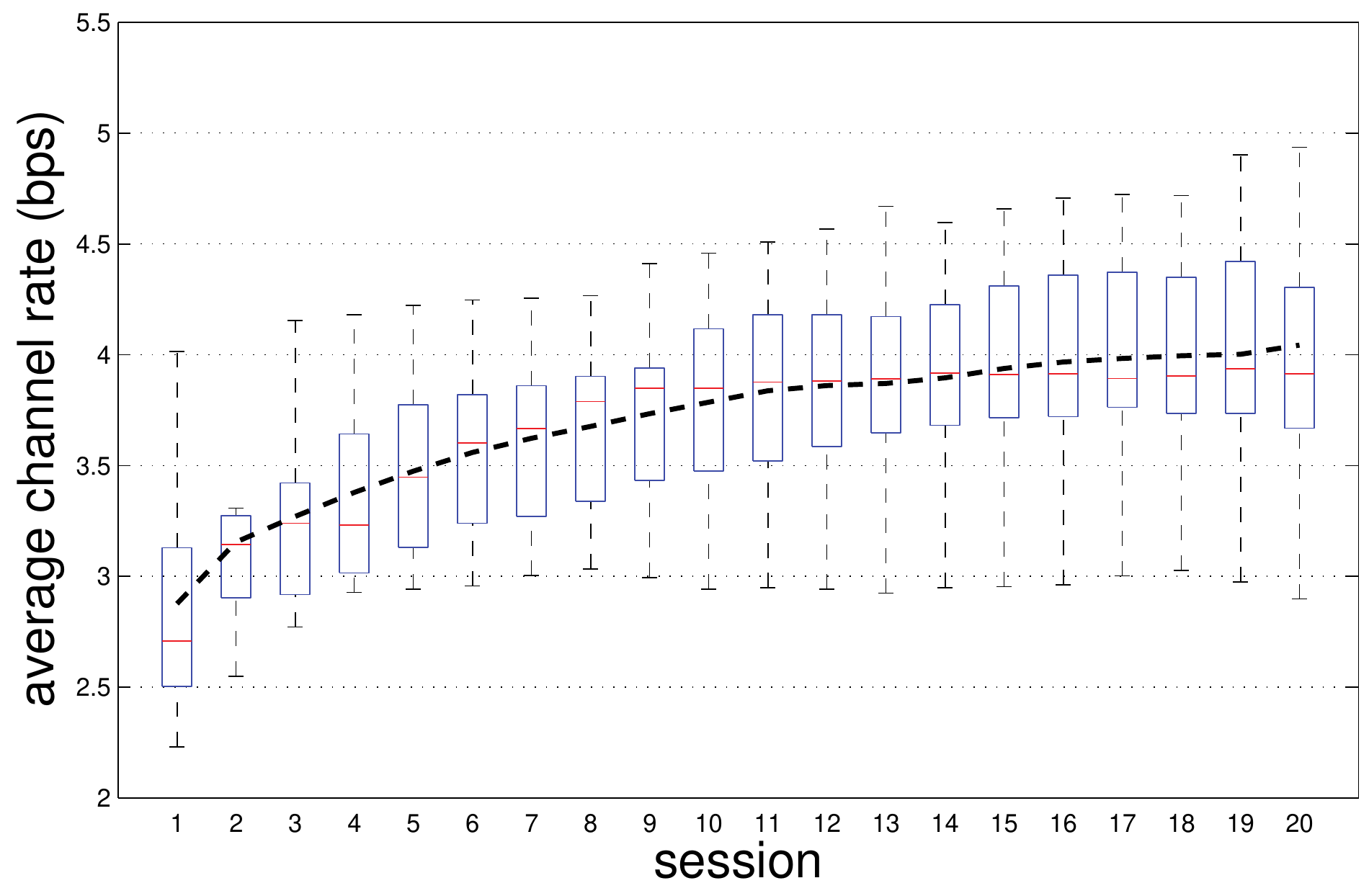}
  }\\
  \subfloat[$\RFixed$]{
    \label{fig:channel_rate_fixed}
    \includegraphics[width=0.75\columnwidth]{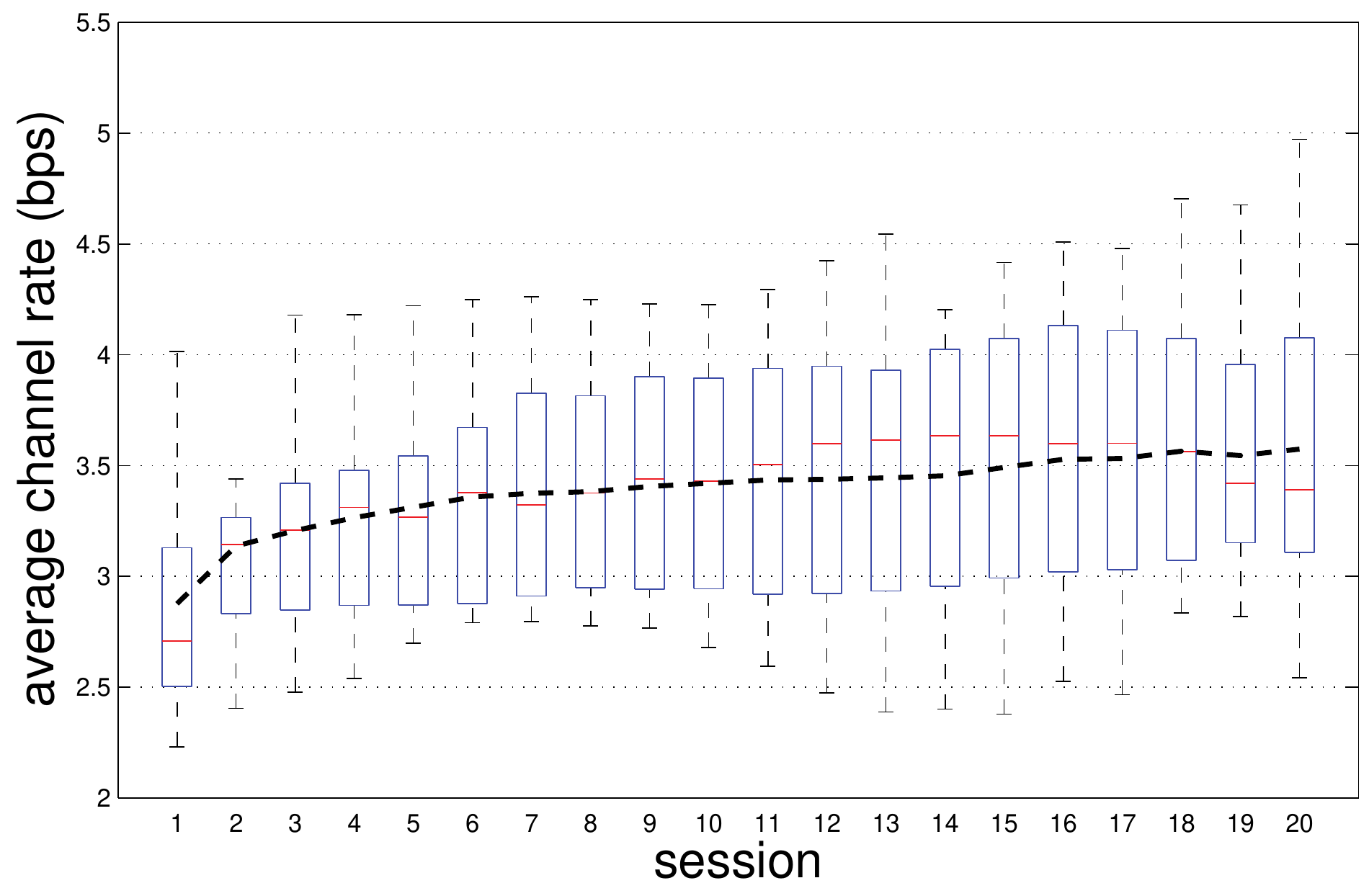}
  }\\
  \subfloat[$\RAdapt - \RFixed$]{
    \label{fig:channel_rate_diff}
    \includegraphics[width=0.75\columnwidth]{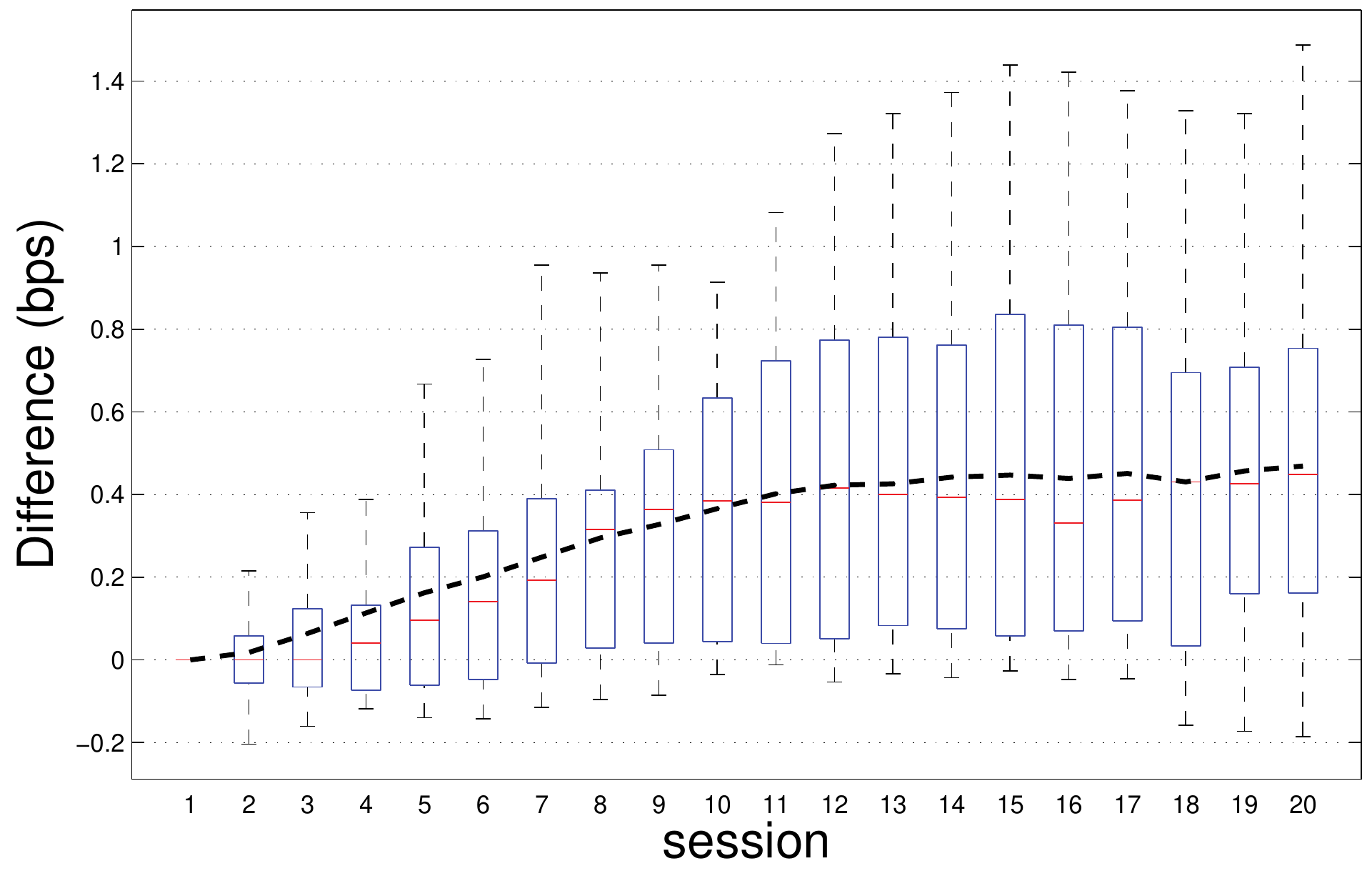}
  }\\
  \caption{Channel rate per session of all users with
    (\ref{fig:channel_rate_adapt}) and without (\ref{fig:channel_rate_fixed})
    presence of machine learning. }
  \label{fig:channel_rate}
\end{figure}

The average channel rates per session of the two conditions $\RAdapt$
and $\RFixed$ are shown in Figure~\ref{fig:channel_rate_adapt} and
Figure~\ref{fig:channel_rate_fixed} respectively. In both conditions,
the results show increases of the channel rate over time where the
improvement in the early sessions seems to be larger than in the later
sessions. Figure~\ref{fig:channel_rate_diff} shows the difference of
$\RAdapt$ and $\RFixed$ which corresponds to the channel rate of the
system when we ignore the effect of user adaptation. From the result,
we observe that the impact of machine adaptation tapers off after 10
sessions.

Although the prototype set was not changing in $\RFixed$, we observe
that channel rate increases over the sessions. To quantify our
confidence to this increase, we perform the paired t-test to compare
the difference between the average channel rate in the first 5
sessions and in the last 5 sessions. We find that the difference is
statistically significant with $p$-value < 0.0011. This suggests that
the users improve the handwriting on their own even without machine
adaptation. In other words, the effect of {\em user adaptation} is
indeed significant.

\begin{figure}[tb]
  \centering
  \subfloat[Writing duration]{
    \label{fig:duration}
    \includegraphics[width=0.75\columnwidth]{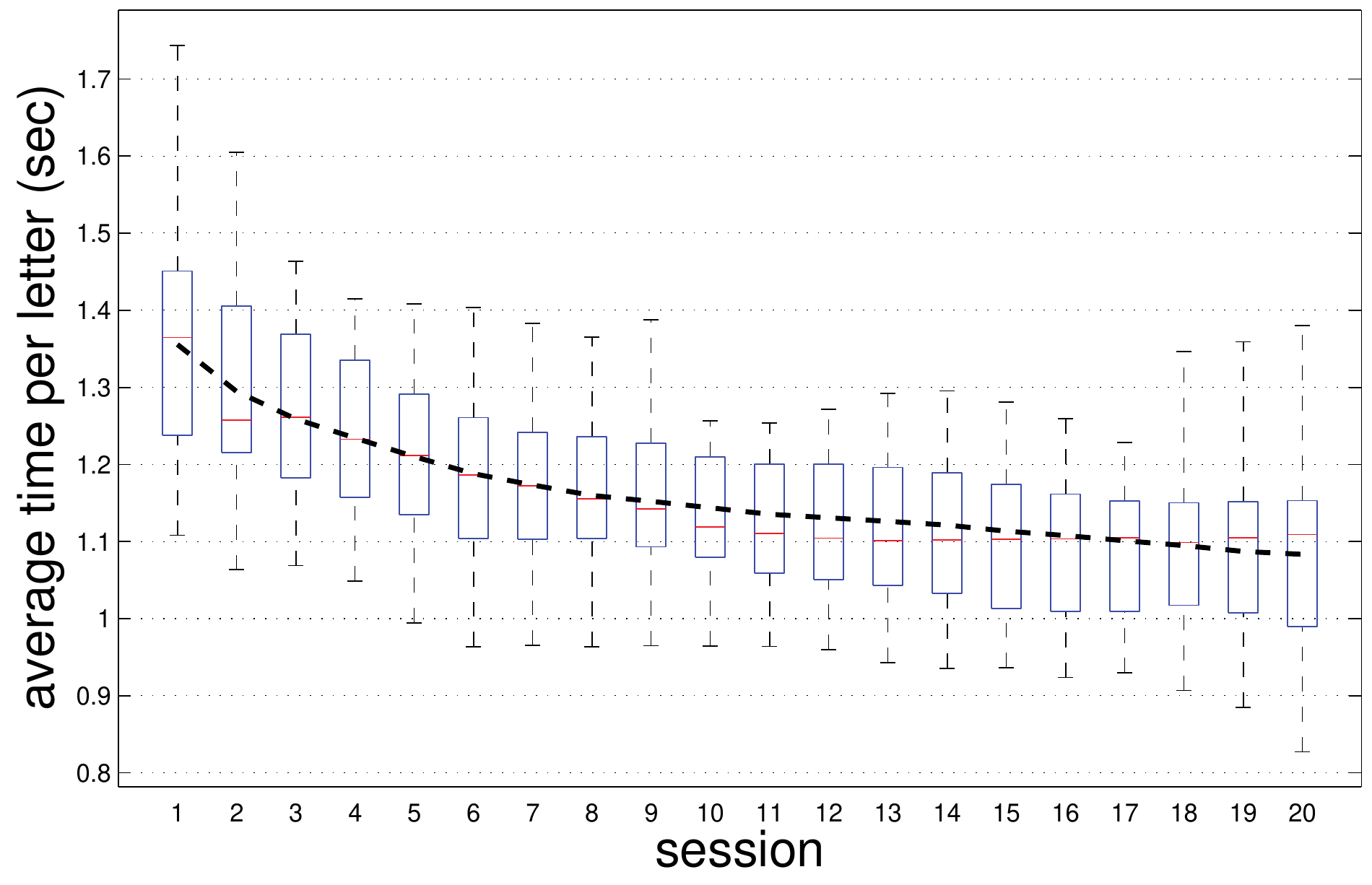}
  }\\
  \subfloat[Mutual information]{
    \label{fig:mutual_information}
    \includegraphics[width=0.75\columnwidth]{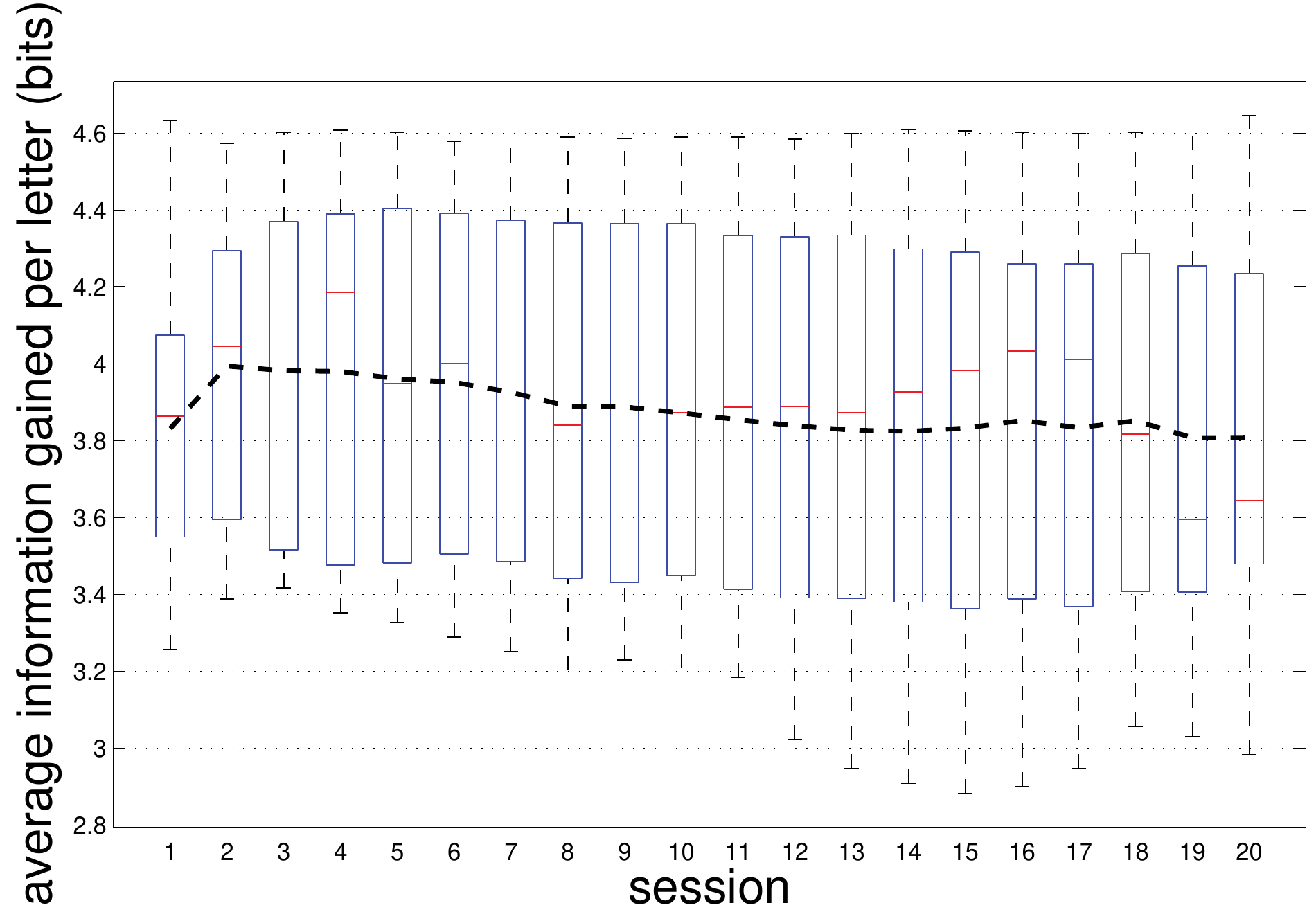}
  }
  \caption{The average writing time per session and the average mutual
    information per session under the condition $\RFixed$.}
\end{figure}

Furthermore, Figure~\ref{fig:duration} and
Figure~\ref{fig:mutual_information} reveal that the major contribution
of {\em user adaptation} comes from the fact that the users write
faster in the last 5 sessions compared to the first 5 sessions ($p <
0.0001$), and not because of the system received more information from
the user ($p = 0.9723$). This result is as expected according to the
law of practice~\cite{Newell1981}.

\begin{figure}[tb]
  \centering
  \subfloat[]{
    \label{fig:channel_rate_per_user}  
    \includegraphics[width=0.75\columnwidth]{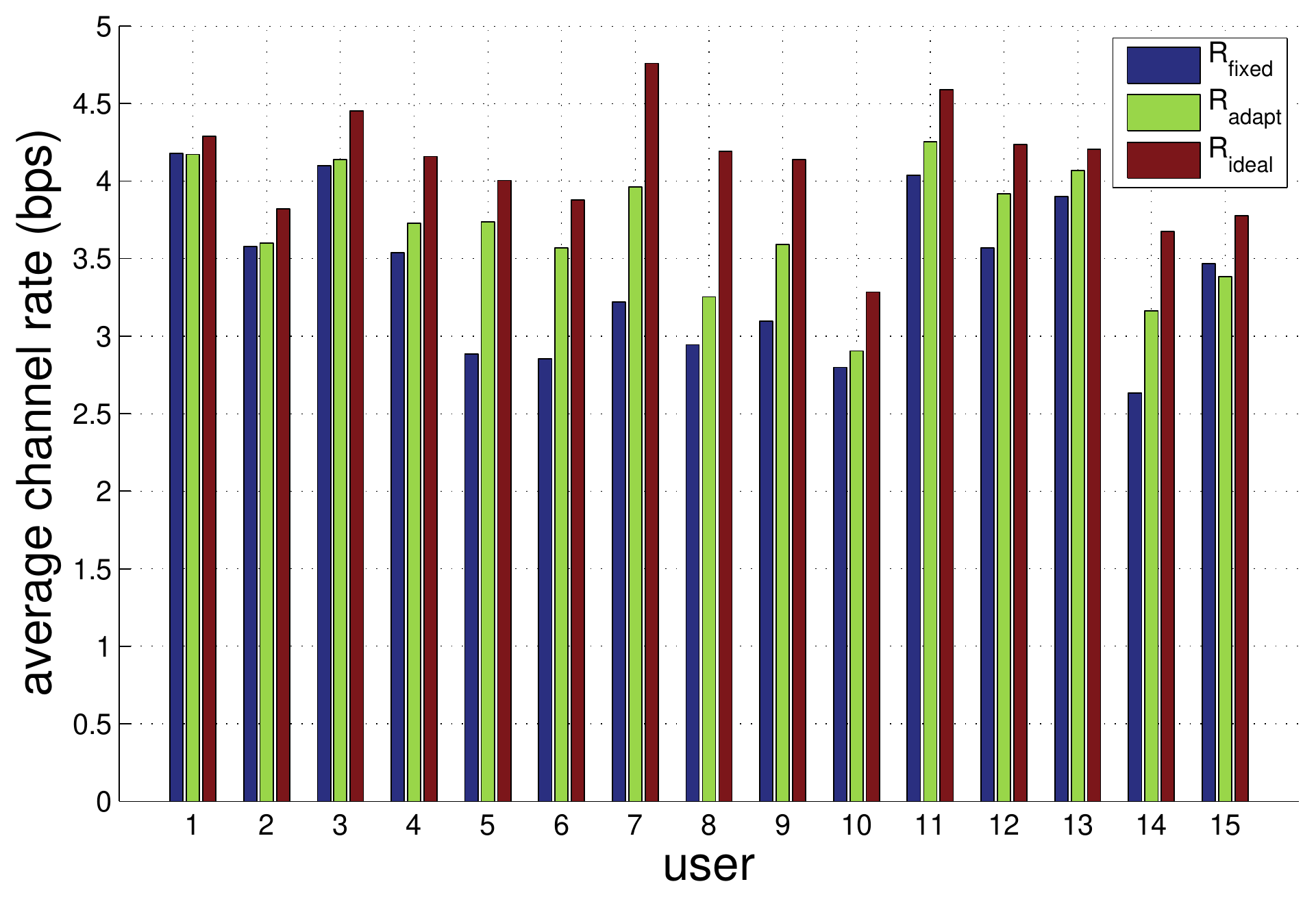}   
  }\\
  \subfloat[]{
    \label{fig:histogram_english_rate} 
    \includegraphics[width=0.75\columnwidth]{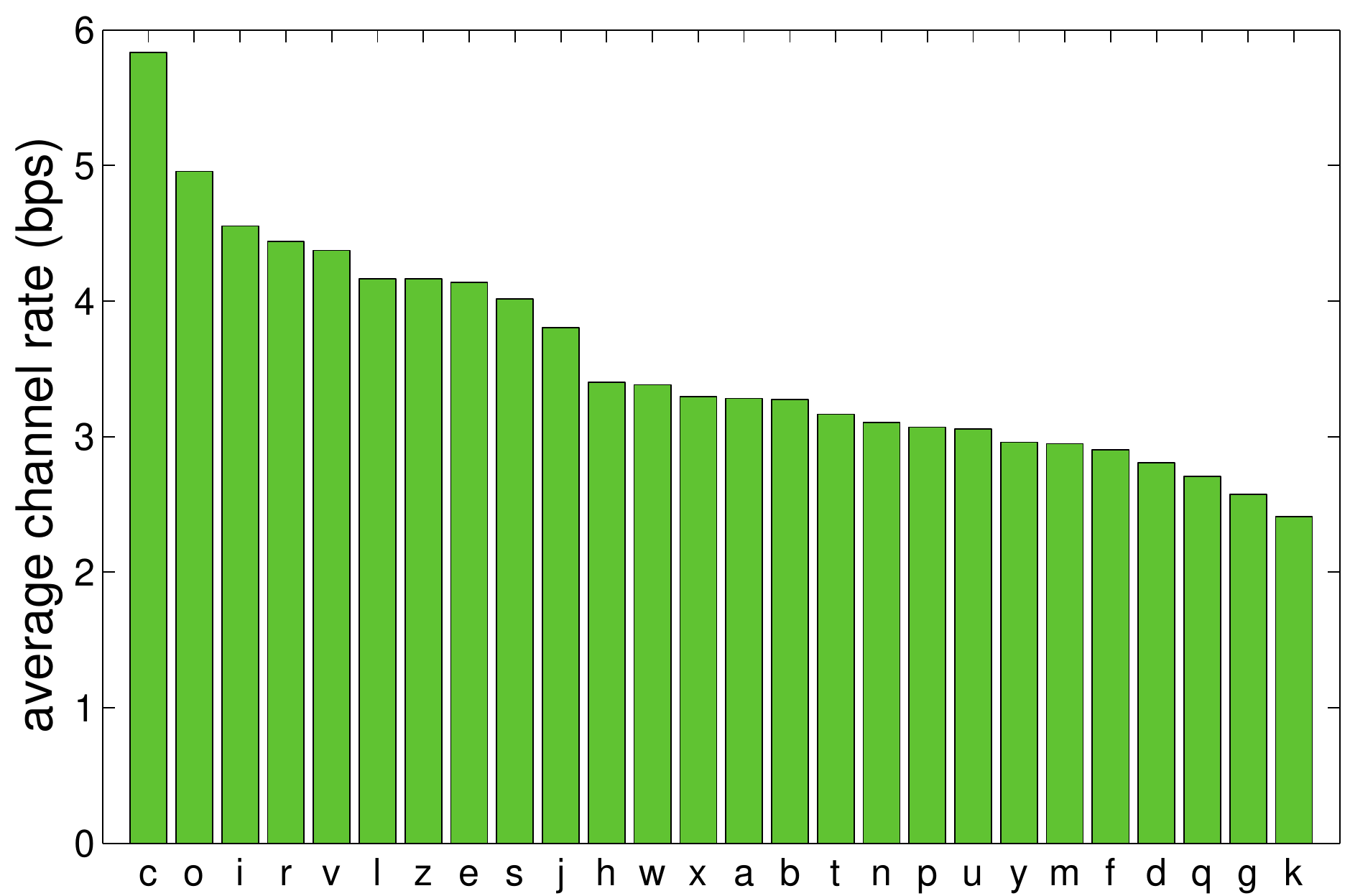}  
  }
  \caption{(a) The average channel rate of each user in $\RAdapt$ and
    $\RFixed$. $\RIdeal$ shows the maximum channel rate possible given
    the average writing speed of each user. (b) Average channel rate
    of each character under the condition $\RAdapt$.}
\end{figure}

We also perform per-user analysis of the channel rate. In
Figure~\ref{fig:channel_rate_per_user}, we compare $\RAdapt$ and
$\RFixed$ for each user. We find that the channel rate of $\RAdapt$
is significantly higher than that of $\RFixed$ with $p < 0.0006$.
This result confirms that the machine adaptation helps improving the
overall channel rate. In addition, we calculate the theoretical
maximum of the channel rate under the assumption of the perfect
recognition, denoted by $\RIdeal$. The maximum rates are given by
$H(\predFinal) / \expectedDuration$ and we approximated $H(\predFinal)
= \log_2(26)$.

In the case of perfect recognition, a simple way to increase the
channel rate is to expand the character set $\intentSet$ to incluse
more symbols. However, in reality, doing so can lead to a recognition
error rate which impairs the channel rate. An interesting future
direction is to design a character set that would maximize the channel
rate. Figure~\ref{fig:histogram_english_rate} reveals the efficiency
of each letter for our handwriting channel. Characters with complex
stokes, such as 'q', 'g','k', are not as efficient as characters with
simple strokes such as 'c' ,'o', 'l'. While this finding is not
surprising, it implies that, for a handwriting system to be truly
efficient, it must allow the user to write in a less complex style
while not losing recognition accuracy. How to exactly design such
system is still an open problem and requires a more elaborate study.

\section{Conclusions}
\label{sec:conclusions}
We presented a information-theoretic framework for quantifying the
information rate of a system that combines a human writer with a
handwriting recognition system. Using the notion of channel rate, we
investigated the impact of machine adaptation and human adaptation in
an adaptive handwriting recognition system. We analyzed data collected
from a small deployment of our adaptive handwriting recognition system
and concluded that both machine adaptation human adaptation have
significant impact on the channel rate.  This result led us to believe
that, for a handwriting recognition system to achieve the maximum
channel rate, both machine adaptation and human adaptation are
required and must be present together. Specifically, such system
should be able to adapt to the user and, at the same time, allow the
users to write or scribble using simple hand movement as improving
writing speed is crucial for attaining a higher channel
rate. Additionally, the system should have a mechanism to giving
feedback to the user when their handwriting cannot be recognized.

\bibliographystyle{elsarticle-num}
\bibliography{library_no_url}

\end{document}